\documentclass[12pt]{article}
\usepackage{epsfig,amssymb,euscript,bbold}
\usepackage{amsmath}
\def\bea{\begin{eqnarray}}
\def\eea{\end{eqnarray}}
\def\be{\begin{equation}}
\def\ee{\end{equation}}

\def\p{\partial}
\def\bE{\bf E}

\def\mathe{\mathfrak{e}}
\numberwithin{equation}{section}
\begin{document}
\title{
Supersymmetric Electromagnetic Waves \\
on Giants and Dual-Giants}
\author{Sujay K. Ashok and Nemani V. Suryanarayana \\
{}\\
{\small{\it Institute of Mathematical Sciences}}\\
{\small{\it C.I.T Campus, Taramani}}\\
{\small{\it Chennai 600113, India.}} \\
\\
{\small{E-mail: {\tt sashok, nemani@imsc.res.in}}} \\
}
\date{}

\maketitle
\begin{abstract}
We set up the BPS equations for a D3-brane moving in $AdS_5 \times S^5$ which preserves two supercharges and with all bosonic fields turned on in the world-volume theory. By solving these, we find generalizations of Mikhailov giants and wobbling dual-giants that include electromagnetic waves propagating on their world-volume. For these giants (dual-giants) we show that the BPS field strength is the real part of the pull-back of a holomorphic 2-form in the ambient space $\mathbb{C}^3$ ($\mathbb{C}^{1,2}$) onto the world-volume. 
\end{abstract}

\newpage

\tableofcontents
\section{Introduction}

Supersymmetric states have played a crucial role in the development of string theories by helping to uncover and substantiate important aspects about dualities. This is especially true of the strong-weak dualities, since many properties of supersymmetric states are protected and therefore remain invariant on both sides of the duality. In the context of the celebrated AdS/CFT correspondence a significant class of BPS states consists of the so-called giant graviton states of type IIB string theory on $AdS_5 \times S^5$. These are supposed to be dual to the R-charged BPS states of the ${\cal N}=4$ $SU(N)$ SYM on $S^3$ under the AdS/CFT dictionary. Such giant gravitons are described classically as solutions of the world-volume theory of a D3-brane propagating in the $AdS_5 \times S^5$ background. They leave unbroken some fraction of the supersymmetries of the background and carry various angular momenta on both $AdS_5$ and $S^5$ parts of the background.

These giant gravitons fall into two classes dependent on whether the D3-brane is point-like in the $AdS_5$ or the $S^5$ part of the geometry. The former are referred to as giants while the latter are called dual-giants. The initial examples of giants \cite{McGreevy:2000cw} (respectively dual-giants \cite{Grisaru:2000zn, Hashimoto:2000zp}) were D3-branes wrapping an $S^3 \subset S^5$ ($AdS_5$) and rotating along the maximal circles of the $S^5$. They carried one angular momentum on $S^5$ and preserved a half of the bulk supersymmetries. Soon it was realized by Mikhailov that there is an infinite class of giants, generalizing the $S^3 \subset S^5$ to a 3-manifold obtained as the intersection of the zero set of a holomorphic function in ${\mathbb C}^3$ with $S^5$ \cite{Mikhailov:2000ya}. Mikhailov giants preserved at least 1/8th of the bulk supersymmetries and generically carried all three independent angular momenta possible on $S^5$. Later on a more general construction of possible giants and dual-giants has been carried out \cite{Kim:2006he, ashoknemani}. 

The solutions in \cite{Mikhailov:2000ya, Kim:2006he, ashoknemani} were obtained by solving the world-volume equations involving only the transverse scalars.  However, the massless content of a D3-brane world-volume consists of transverse scalars, a $U(1)$ gauge field and fermionic fields. In general, one expects the existence of giant and dual-giant like solutions with all these fields turned on. In fact the authors of \cite{Kim:2005mw, Sinha:2007ni} were able to provide the first, albeit isolated, examples of giants and dual-giants with non-trivial world-volume electromagnetic fields. One would like to know if more general classes of such solutions exist.

In this paper we consider the problem of finding {\it all} giant and dual-giant like D3-brane solutions with all the massless bosonic fields (namely, the scalars and the $U(1)$ gauge field) on the world-volume turned on. We use the techniques employed in \cite{ashoknemani} to set up and solve the BPS equations of the D3-brane world-volume theory involving all the massless bosonic fields demanding that at least two given supersymmetries of the bulk are preserved.

Our solutions can be elegantly described using the auxiliary space ${\mathbb C}^{1,2} \times {\mathbb C}^3$ with coordinates $\{\Phi^0, \Phi^1, \Phi^2; Z_1, Z_2, Z_3\}$ where the $AdS_5 \times S^5$ can be embedded as ${|\Phi^0|}^2 - {|\Phi^1|}^2 - {|\Phi^2|}^2 = l^2$ and $|Z_1|^2 + |Z_2|^2 + |Z_3|^2 = l^2$. The BPS equations, it turns out, fall into two classes: those that are independent of the electromagnetic field strength $F$ and those that depend on $F$. The $F$-independent equations describe the embedding geometry of the D3-brane and coincide with the equations derived in \cite{ashoknemani}. Let us briefly recall the particular solutions to these equations which are relevant for the present work. These are two classes of D3 branes that preserve four out of the thirty two bulk supersymmetries (see \cite{ashoknemani} for details).

\vspace{.5cm}
\noindent\underline{Wobbling dual-giants}: Defining $Y^i = \Phi^i Z_1$, these are described as the intersection of 
\begin{align}
f(Y^0, Y^1, Y^2) &=0\quad\text{and}\cr 
Z_2 = Z_3 &= 0  
\end{align}
with $ AdS_5 \times S^5$. 

\vspace{.5cm}
\noindent\underline{Mikhailov giants}: These are the intersections of $AdS_5 \times S^5$ with 
\begin{align}
f(X^1, X^2, X^3) &= 0\quad\text{and}\cr 
\Phi^1 = \Phi^2 &= 0\,,
\end{align}
where $X^i = Z_i \Phi^0$. 

In order to solve the BPS equations involving the field strength we make an ansatz, namely, that it be the pull-back of a general spacetime 2-form
\be\label{Fansatz}
F =\text{P} \left[ \sum_{a,b} \chi_{ab} \, e^{ab}\right] \,,
\ee
where the $e^{a}$ are some basis of vielbein in spacetime and the $\chi_{ab}$ are arbitrary functions in spacetime. We find that given such an ansatz, solving the BPS equations, the equations of motion and the Bianchi identity is equivalent to finding a specific complex 2-form $G$ in spacetime which is closed when pulled back onto the world-volume. The field strength of the supersymmetric world-volume gauge field configuration is then given by the real part of the pull-back of $G$ which, for the D-branes under consideration, takes the form 
\be
G = \sum_{ij=0,1,2}G_{ij}\, (Y) dY^i \wedge\, dY^j \,.
\ee
For the giants, a similar ansatz and consequent  analysis leads to the conclusion that the world-volume gauge field strength is again given by the real part of 
\be
\tilde G = \sum_{ij=1,2,3} \tilde G_{ij}(X)\, dX^i \wedge\, dX^j \,,
\ee
pulled back onto the world-volume.

In section 2, we present the BPS equations for supersymmetric configurations of D3 branes with non-trivial electromagnetic fluxes on the world-volume. In section 3, we use the ansatz \eqref{Fansatz} and solve the algebraic BPS equations for the dual-giants. We then use the ansatz to combine the Bianchi identity and equations of motion of the gauge field into a single constraint, the closure of  a complex 2-form (whose real part is $F$) when pulled back onto the D-brane world-volume. In section 4, we repeat the analysis for giants. We then test our general results in section 5, by solving the case of the spherical (dual-) giant and recovering the existing solutions in the literature. We conclude with a discussion of some open issues and future work. Some technical details are collected in the appendices.  

\section{Supersymmetry analysis}

We begin by studying the kappa-projection conditions that ensure supersymmetry for a D3-brane embedded in $AdS_5\times S^5$ with world-volume gauge field flux $F$ turned on. We follow the conventions of \cite{ashoknemani}. The metric on $AdS_5 \times S^5$ written in global coordinates is 
\begin{multline}
\frac{ds^2}{l^2} = -\cosh^2\rho \, d\phi_0^2 + d\rho^2+\sinh^2\rho \, (d\theta^2+\cos^2\theta \, d\phi_1^2+\sin^2\theta \, d\phi_2^2)\cr
+ d\alpha^2 + \sin^2\alpha \, d\xi_1^2+\cos^2\alpha \, (d\beta^2 +\sin^2\beta \, d\xi_2^2+\cos^2\beta \, d\xi_3^2)
\end{multline}
where $\phi_0 = \frac{t}{l}$. We choose the following frame for the $AdS_5$ part of the metric
\begin{align}\label{adsframe}
e^0 &= l [\cosh^2\rho \, d \phi_0 - \sinh^2\rho \, (\cos^2\theta d\phi_1 + \sin^2\theta d \phi_2)], \cr 
e^1 &= l \, d\rho\,,\qquad e^2 = l \sinh \rho \, d\theta, \cr
e^3 &= l \cosh\rho \sinh\rho \, ( \cos^2\theta ~ d\phi_{01} + \sin^2\theta  ~ d\phi_{02})\, \cr
e^4 &= l \sinh\rho \, \cos\theta \sin\theta ~ d\phi_{12}
\end{align}
where $\phi_{ij} = \phi_i - \phi_j$. For the $S^5$ part we choose the frame
\begin{align}\label{sframe}
e^5 &= l \, d\alpha, \qquad e^6 = l \, \cos\alpha \, d\beta, \cr
e^7 &= l \, \cos\alpha \sin\alpha \, (\sin^2\beta \, d\xi_{12} + \cos^2\beta \, d\xi_{13}), \cr
e^8 &= l \, \cos\alpha \cos\beta \sin\beta \, d\xi_{23}, \cr
e^9 &=  l \, (\sin^2\alpha \, d\xi_1 + \cos^2\alpha \sin^2\beta \, d\xi_2 +  \cos^2\alpha \cos^2\beta \,d\xi_3)
\end{align}
where $\xi_{ij} = \xi_i - \xi_j$. The Killing spinor for the $AdS_5 \times S^5$ background adapted to the above frame  is given by   
\begin{multline}
\label{adskss2}
\epsilon = ~ e^{-\frac{1}{2} (\Gamma_{79} - i \Gamma_5 \, \tilde \gamma) \, \alpha}
e^{-\frac{1}{2} (\Gamma_{89} - i \Gamma_6 \tilde \gamma) \beta} \, e^{\frac{1}{2} 
\xi_1 \Gamma_{57}} \, e^{\frac{1}{2} \xi_2  \Gamma_{68}} \,
e^{\frac{i}{2} \xi_3  \Gamma_9 \, \tilde \gamma} \cr
\times e^{\frac{1}{2} \rho \, (\Gamma_{03} + i \Gamma_1 \,
\gamma)} \,  e^{\frac{1}{2}
\theta \, (\Gamma_{12} + \Gamma_{34})} \, e^{\frac{i}{2} \phi_0 \, \Gamma_0 \, \gamma} \, e^{-\frac{1}{2} \phi_1 \Gamma_{13}} \,
e^{-\frac{1}{2} \phi_2 \Gamma_{24}} \, \epsilon_0
\end{multline}
where $\epsilon_0$ is an arbitrary 32-component Weyl spinor satisfying $\Gamma_0 \cdots \Gamma_9 \epsilon_0 = - \epsilon_0$ and $\gamma = \Gamma^{01234}$ and $\tilde \gamma = \Gamma^{56789}$.

In the rest of this section, we seek the full set of BPS equations for a D3-brane in $AdS_5 \times S^5$ with non-trivial world-volume gauge fluxes, and preserves two supersymmetries out of thirty two. Clearly this choice of preserved supersymmetries is non-unique. Without loss of generality we choose them to be the ones obtained in \cite{Mandal:2006tk}. So we take superymmetries preserved by the D3-brane to be those that survive the following projections
\begin{equation}
\label{projections}
\Gamma_{57} \epsilon_0 = \Gamma_{68} \epsilon_0 = i \epsilon_0\,, \qquad \Gamma_{09} \epsilon_0 =  -\epsilon_0\,,\qquad \Gamma_{13} \epsilon_0 = \Gamma_{24} \epsilon_0 = -i \epsilon_0 \,. 
\end{equation}
With these projections the Killing spinor simplifies to
\begin{equation}
\label{reducedks}
\epsilon = e^{\frac{i}{2} (\phi_0 + \phi_1 + \phi_2 + \xi_1 + \xi_2 + \xi_3) } \epsilon_0.
\end{equation}
From now on, we set $l=1$ for convenience. Next we obtain the equations that any D3-brane should satisfy to preserve (at least) these two supersymmetries. 

The ansatz we take for the D3-brane is the most general one, such that all spacetime coordinates $(r, \theta, \phi_0, \phi_1, \phi_2, \alpha, \beta, \xi_1, \xi_2, \xi_3)$ are functions of the world-volume coordinates $(\sigma_0, \sigma_1, \sigma_2, \sigma_3)$. The world-volume gamma matrices are
\begin{equation}
\gamma_i = {\mathfrak e}^a_i \Gamma_a
\end{equation}
where the bold-face symbol ${\mathfrak e}^a_i = e^a_\mu \partial_i X^\mu$ with $i \in \{\sigma_0, \sigma_1, \sigma_2, \sigma_3\}$ is the pull-back of $e^a_\mu$ onto the world-volume. 
The kappa symmetry equation in the presence of world-volume fluxes is given by \cite{Cederwall:1996pv, Bergshoeff:1996tu}
\be
\label{kappaproj}
\epsilon^{ijkl} 
\left[ \frac{1}{4!}\gamma_{ijkl} I +\frac{1}{4}F_{ij}\gamma_{kl} I K+
 \frac{1}{8}F_{ij}F_{kl}I \right] \epsilon = 
\sqrt{-\det (h+F)} \, \epsilon \,,
\ee
where the operators $K$ and $I$ are defined so that
\be
K\epsilon = \epsilon^* \qquad \hbox{and} \qquad I \epsilon = -i \epsilon \,.
\ee
We would like to preserve the same supersymmetries as for the giant gravitons without electromagnetic flux. This means we impose the condition 
\begin{equation}
\label{BPScond1}
 \gamma_{\sigma_0\sigma_1\sigma_2\sigma_3} \epsilon = i \, \sqrt{-\det h} \, \epsilon 
\end{equation}
where $\gamma_{\sigma_0\sigma_1\sigma_2\sigma_3} = \mathfrak{e}^a_0 \mathfrak{e}^b_1
\mathfrak{e}^c_2 \mathfrak{e}^d_3 \Gamma_{abcd}$. 
This leads to the constraint 
\begin{equation}
\label{BPScond2}
\epsilon^{ijkl}F_{ij}\gamma_{kl}\epsilon_0^{*} = 0\,. 
\end{equation}
Substituting (\ref{BPScond1}, \ref{BPScond2}) into (\ref{kappaproj}) we get the further condition
\begin{equation}
\label{BPScond3}
\sqrt{-\det h} + {\rm Pf}[F]  = \sqrt{-\det(h+F)} \,,
\end{equation}
where ${\rm Pf}[F] = \frac{1}{8} \epsilon^{ijkl} F_{ij} F_{kl}$ which we sometimes denote  by ``$F \wedge F$''. The BPS equations that follow from (\ref{BPScond1}, \ref{BPScond2}, \ref{BPScond3}) are most compactly expressed in terms of the pull-back of the bulk complex 1-forms
\be\label{cplxforms}
E^1 = e^1 -i e^3 \qquad E^2 = e^2 -i e^4 \qquad E^5 = e^5 +i e^7 \qquad E^6 = e^6 +i e^8 \,, 
\ee
and the real 1-forms
\be
E^0 = e^0 + e^9 \qquad \text{and}\qquad E^{\bar 0} = e^0 - e^9 \, . 
\ee
It can be shown that the equation (\ref{BPScond2}) gives rise to the following conditions:
\begin{align}\label{bpsfeqns}
F\wedge {\bf E}^{AB} &= 0 \quad\text{for}\quad A,B = \{1,2,5,6\} \cr
F\wedge {\bf E}^{0}\wedge {\bf E}^{A} &= 0 \quad\text{for}\quad A = \{1,2,5,6\} \cr
F\wedge (\mathe^{09}+i \Omega) &= 0\,, \quad 
\end{align}
where $\Omega = \tilde{\omega}-\omega$, with
\begin{align}\label{kahlerforms}
\tilde \omega &= \mathfrak{e}^{13} + \mathfrak{e}^{24} = - \frac{i}{2} ( {\bf E}^{1\bar{1}} +  {\bf E}^{2 \bar{2}}) \quad\text{and}\cr
\omega &= \mathfrak{e}^{57} + \mathfrak{e}^{68} = ~~\frac{i}{2} ( {\bf E}^{5\bar{5}} +  {\bf E}^{6 \bar{6}}) \, .
\end{align}
In these equations, by $F \wedge {\bf E}^{ab}$ we mean $\frac{1}{2} \epsilon^{ijkl} F_{ij} {\bf E}^a_k {\bf E}^b_l$. As for the real 1-forms, the boldfaced characters refer to the pull-back of the forms onto the world-volume. Next we would like to solve (\ref{BPScond3}). For this we note the following identity 
\begin{multline}
-\det (h+F) =  -\det h - ({\rm Pf}[F])^2 + (\mathfrak{e}^{09} \wedge F)^2 \cr
- \sum_{A=1, 2, 5, 6} \left[|\mathfrak{e}^9 \wedge {\bf E}^A \wedge F|^2 - |\mathfrak{e}^0 \wedge {\bf E}^A\wedge F|^2 \right]  \cr-\sum_{A<B} |{\bf E}^{AB} \wedge F|^2 - (\Omega \wedge F)^2 + (\Omega \wedge \Omega) \, {\rm Pf} [F] \, .
\end{multline}
We provide some details in Appendix A on how to derive this expression. Substituting the BPS conditions (\ref{bpsfeqns}) linear in the field strength into this expression, and noting that $\Omega \wedge \Omega = \frac{1}{2} (\tilde \omega - \omega) \wedge (\tilde \omega - \omega) = 0$ for time-like D3-branes (see \cite{ashoknemani} for details) we obtain
\begin{equation}
\label{detidentty}
\det (h+F) = \det h + (F\wedge F)^2 \, .
\end{equation}
Demanding the consistency of  (\ref{BPScond3}) and (\ref{detidentty}) we get the last of the $F$- dependent BPS conditions
\begin{equation}
\label{BPScond4}
F \wedge F = 0 \,. 
\end{equation}
This in turn implies $\det (h+F) = \det h$ for the BPS configurations we seek. Finally the BPS constraints that do not involve the field strength are the same as those obtained in \cite{ashoknemani}:
\begin{align}
{\bf E}^{ABCD} &= 0, ~~~ {\bf E}^{0 ABC} =0 \cr
(\mathfrak{e}^{09} + i \, \Omega) \wedge {\bf E}^{AB} &= 0 ~~{\rm for} ~~ A, B, C, D = 1, 2, 5, 6 \cr
\Omega \wedge \Omega &=0 \, .
\end{align}
for time-like brane embeddings. In these equations we understand ${\bf E}^{abcd}$ to be the function (0-form) $\epsilon^{ijkl} {\bf E}^a_i {\bf E}^b_j {\bf E}^c_k {\bf E}^d_l$. Using all the BPS conditions, for a time-like D3-brane one obtains
\be\label{dethdgg}
\sqrt{- \det h} = e^{09} \wedge \Omega  =  i \, {\bf E}^{0 \bar 0} \wedge \sum_A {\bf E}^{A \bar A} \,.
\ee
For dual-giants, the BPS equations take the simplified form 
\begin{align}\label{bpseqnsdgg}
{\bf E}^{0\bar{0} 12} &= 0, \cr
{\bf E}^0 \wedge \{ {\bf E}^1, {\bf E}^2 \} \wedge \tilde {\bf \omega} &= 0, \cr
{\bf E}^5 = {\bf E}^6 &= 0\,,
\end{align}
while for giants, they take the form
\begin{align}\label{bpsqnsgg}
{\bf E}^{0\bar{0} 56} &= 0, \cr
{\bf E}^0 \wedge \{ {\bf E}^5, {\bf E}^6 \} \wedge \omega &= 0, \cr
{\bf E}^1 = {\bf E}^2 &= 0\,.
\end{align}
We now restrict our attention to dual-giant gravitons. Using the fact that the field-strength $F$ is real, the F-dependent BPS conditions for the dual-giants take the form
\begin{subequations}\label{Fdualgiants}
\begin{align}
F\wedge {\bf E}^{0\bar{0}} &= 0 \\ 
F\wedge {\bf E}^0 \wedge \{{\bf E}^1, {\bf E}^2, {\bf E}^{\bar 1}, {\bf E}^{\bar 2} \} &= 0 \\
F\wedge F &= 0\\
F\wedge  ( {\bf E}^{1\bar{1}} +  {\bf E}^{2 \bar{2}}) &= 0 \\
F\wedge \{ {\bf E}^{12}, {\bf E}^{{\bar 1}{\bar 2}} \} &= 0 \, .
\end{align}
\end{subequations}
Next, we turn to solving these equations.

\section{General dual-giant solutions}

Apart from the algebraic conditions (\ref{Fdualgiants}) on $F$ there are also differential conditions, namely, its equations of motion and the Bianchi identity. Let us first solve the algebraic conditions (\ref{Fdualgiants}). At this point we make an important assumption, that the field strength $F$ on the world-volume can be written as a pull-back of a space-time 2-form onto the world-volume. This assumption allows us to solve the above algebraic conditions in a rather straightforward way. There are fifteen 2-forms that can be constructed out of the six bulk 1-forms of relevance $\{ {\bf E}^0, {\bf E}^{\bar 0}, {\bf E}^1, {\bf E}^2, {\bf E}^{\bar 1}, {\bf E}^{\bar 2} \}$ and the 2-form we seek is a real linear combination of all these two-forms. 

We start by assuming the most general ansatz for $F$:
\begin{equation}
F = {\rm Re} \left[ \chi_{0\bar 0} {\bf E}^{0 \bar 0} + \sum_A (\chi_{0A} {\bf E}^{0A} + \chi_{\bar 0 A} {\bf E}^{\bar 0 A}) + \sum_{A \le B} (\chi_{AB} {\bf E}^{AB} + \chi_{A \bar B} {\bf E}^{A \bar B}) \right]
\end{equation}
and then substitute it back into \eqref{Fdualgiants}. After using the BPS equations \eqref{bpseqnsdgg} one will still be left with linear combinations of nine of the non-vanishing 4-forms on the left hand side of \eqref{Fdualgiants}. We treat these nine 4-forms to be independent and set their coefficients to zero. This is justified because of our earlier assumption that $F$ can be written as the pull-back of a space-time 2-form irrespective of the details of the world-volume embedding equations. With this assumption it can be shown that the algebraic conditions \eqref{Fdualgiants} can be solved if and only if 
\begin{equation}\label{Fanswer}
F = \rm{Re}[\chi_{01} {\bf E}^{01}+ \chi_{02} {\bf E}^{02} + \chi_{12} {\bf E}^{12}]
\end{equation}
where $\chi_{01}, \chi_{02}, \chi_{12}$ are arbitrary complex functions of the bulk coordinates restricted to the world-volume.

\subsection{Bianchi identity and the equations of motion}

It now remains to solve for the parameters $\{\chi_{01}, \chi_{02}, \chi_{12}\}$ by imposing the Bianchi identity and the equation of motion for the gauge field. These are
\be\label{BIandEOM}
dF = 0 \qquad\text{and}\qquad \p_{i}X^{ij} =0 \,,
\ee
where 
\be
\label{defX}
X^{ij} =  \frac{1}{2}\sqrt{-\det(h+F)}\big[ (h+F)^{-1}-(h-F)^{-1}\big]^{ij} \,.
\ee
For any $4 
\times 4$ symmetric matrix $h$ whose components can be written as $h_{ij} = \mathfrak{e}^a_i \mathfrak{e}^b_j \eta_{ab}$ and for any antisymmetric $4 \times 4$ matrix $F$, we note the identity 
\begin{multline}
\det (h+F) [ (h+F)^{-1} - (h-F)^{-1}]^{ij} \\ = - (\frac{1}{4} \epsilon^{pqrs}F_{pq}F_{rs}) \, \epsilon^{ijmn} F_{mn}   - (\frac{1}{2} \epsilon^{pqrs}F_{pq} \mathfrak{e}^a_r \mathfrak{e}^b_s) \, \eta_{ac}\eta_{bd} \epsilon^{ijmn} \mathfrak{e}^c_m \mathfrak{e}^d_n \,.
\end{multline}
This comes about by recognizing the left hand side of this equation as the difference of Adjoint matrices of $(h+F)$ and $(h-F)$, which, in turn, follows from the fact that $\det(h+F)=\det(h-F)$. 
Given the definition of $X^{ij}$ in \eqref{defX} and using the BPS equation $F \wedge F =0$, we obtain
\begin{equation}\label{originalX}
X^{ij} = \frac{1}{2\sqrt{- \det h}} (\frac{1}{2} \epsilon^{pqrs}F_{pq} \mathfrak{e}^a_r \mathfrak{e}^b_s) \, \eta_{ac}\eta_{bd} \epsilon^{ijmn} \, \mathfrak{e}^c_m \mathfrak{e}^d_n\,.
\end{equation}
We will need to simplify this further using the BPS equations. Before proceeding further we note that the equation of motion $\partial_i X^{ij} = 0$ can be written as $d\tilde X = 0$ for the 2-form defined as
\begin{equation}
\tilde{X} = \frac{1}{4}\epsilon_{ijmn}X^{mn} \, d\sigma^i \wedge d\sigma^j.
\end{equation}
We will therefore work with the 2-form $\tilde X$ and simplify it using our ansatz for the field strength $F$ and the BPS equations. Substituting the ansatz we have for $F$ and retaining only those terms which (potentially)  survive after using the BPS equations \eqref{Fdualgiants} one finds
\begin{equation}
\tilde X = - \frac{1}{\sqrt{-\det h}} \left[ (F \wedge {\bf E}^{\bar 0 \bar 1}) \, {\bf E}^{0 1} 
+ (F \wedge {\bf E}^{\bar 0 1}) \, {\bf E}^{0 \bar 1}
+ (F \wedge {\bf E}^{\bar 0 \bar 2}) \, {\bf E}^{0 2}
+ (F \wedge {\bf E}^{\bar 0 2}) \, {\bf E}^{0 \bar 2} \right] \,,
\end{equation}
where we have re-expressed $\mathe^{ab}$ in terms of ${\bE}^{ab}$. We claim that when $F = \rm{Re}[ \chi_{01} \, {\bf E}^{01} + \chi_{02} \, {\bf E}^{02} + \chi_{12} \, {\bf E}^{12}]$, this expression is equal to 
\begin{equation}
\tilde X =  \frac{i}{\sqrt{-\det h}} ({\bf E}^{0 \bar 0 1 \bar 1} + {\bf E}^{0 \bar 0 2 \bar 2}) \, \rm{Im} [ \chi_{01} \, {\bf E}^{01} + \chi_{02} \, {\bf E}^{02} + \chi_{12} \, {\bf E}^{12}]\,.
\end{equation}
This can be shown by using the BPS equations and the identity
\begin{equation}
{\bf E}^{a[bcd} ~ {\bf E}^{f]a} = 0\,,
\end{equation}
where, as before, we understand ${\bf E}^{abcd}$ to mean $\epsilon^{ijkl} {\bf E}^a_i {\bf E}^b_j {\bf E}^c_k {\bf E}^d_l$, and treat ${\bf E}^{ab}$ as the rank-2 anti-symmetric tensor ${\bf E}^a_i {\bf E}^b_j - {\bf E}^a_j {\bf E}^b_i$. Finally restricting \eqref{dethdgg} to the case of dual-giants, we have 
\begin{equation}
\sqrt{-\det h} = i ({\bf E}^{0 \bar 0 1 \bar 1} + {\bf E}^{0 \bar 0 2 \bar 2})\,.
\end{equation}
Using this, we obtain the result
\be\label{finalX}
\tilde{X} = \text{Im}[ \chi_{01} {\bf E}^{01} + \chi_{02} {\bf E}^{02} + \chi_{12} {\bf E}^{12}]\, .
\ee
This is a remarkable simplification, considering the original expression \eqref{defX} we started with, which was highly non-linear in the pull-back of the vielbeins and the field strength $F$. This can be attributed to the effectiveness of the BPS equations in simplifying the problem. 

\subsection{Solving the Bianchi identity and equations of motion}

Our final expressions for the real 2-forms $F$ and $\tilde{X}$ in \eqref{Fanswer} and \eqref{finalX} makes it natural to define a complex 2-form 
\be\label{defG}
G = F + i \tilde{X} =  \chi_{01} {\bf E}^{01} + \chi_{02} {\bf E}^{02} + \chi_{12} {\bf E}^{12} 
\ee
in terms of which the Bianchi identity and the equations of motion can be combined into the single equation 
\be
dG = 0 \,,
\ee
where $dG$ refers to the exterior derivative of $G$ on the world-volume. However, for differential forms, the pull-back operation and the exterior differentiation commute. So, we treat the right hand side of \eqref{defG} as a spacetime 2-form, compute the exterior derivative in spacetime, and then require that the resulting 3-form vanishes, when pulled back onto the world-volume. This is what we will do in the remainder of the section. 

\vskip .5cm
\noindent{\underline{\bf Solving $dG=0$}}
\vskip .5cm
Let us first recall some facts regarding the wobbling dual-giant solution. It is known that a wobbling dual-giant is described by a polynomial equation of the form\footnote{The wobbling dual-giants rotating along other maximal circles can be obtained from these by replacing $Z_1$ by an appropriate phase.}
 \be\label{fiszero}
 f(Y_0, Y_1, Y_2) = 0 \,,
 \ee
where $Y^i = \Phi^i Z_1$ with 
\begin{align}
\Phi^0 &= \cosh \rho \, e^{i\phi_0}\qquad \Phi^1 = \sinh \rho \cos\theta \, e^{i\phi_1}\qquad \Phi^2 = \sinh \rho \sin\theta \, e^{i\phi_2}\cr 
Z_1 &= e^{i\xi_1}\qquad\text{and}\qquad Z_2 = Z_3 = 0 \,. 
\end{align}
On such a $3+1$ dimensional world-volume, we seek a closed complex $2$-form of the kind \eqref{defG}. Since the equation of the D-brane is written purely in terms of the $Y^i$ variables, our search for a closed $2$-form will be greatly simplified by rewriting \eqref{defG} in terms of the differentials $dY^i$. Given the definition of the $Y^i$ above, one can relate the differentials $dY^i$ to the 1-forms $E^A$ and $E^{\bar A}$. Inverting these relations and wedging them together, one finds the following relations:
\begin{align}
\label{EstoYs}
E^{01} &= -i \cosh\rho \sinh\rho \, \frac{dY^0}{Y^0} \wedge \left[\cos^2\theta \, \frac{dY^1}{Y^1} + \sin^2\theta \, \frac{dY^2}{Y^2}\right], \cr
E^{02} &= i\sinh \rho \cos\theta \sin\theta \, \left[ \cosh^2 \rho \, \frac{dY^0}{Y^0} \wedge (\frac{dY^1}{Y^1} - \frac{dY^2}{Y^2}) + \sinh^2 \rho \, \frac{dY^1}{Y^1} \wedge \frac{dY^2}{Y^2} \right], \cr
E^{12} &= \cosh \rho \sinh^2 \rho \cos\theta \sin\theta \, \left[ \frac{dY^0}{Y^0} \wedge (\frac{dY^1}{Y^1} - \frac{dY^2}{Y^2}) + \frac{dY^1}{Y^1} \wedge \frac{dY^2}{Y^2} \right]\,,
\end{align}
from which we can write the complex $2$-form $G$ in the form
\begin{align}
G &= \chi_{01} {\bE}^{01} + \chi_{02} {\bE}^{02} + \chi_{12} {\bE}^{12} \cr
&:=  G_{01} \frac{dY^0}{Y^0} \wedge \frac{dY^1}{Y^1} 
+ G_{02}  \frac{dY^0}{Y^0} \wedge \frac{dY^2}{Y^2} 
+ G_{12} \frac{dY^1}{Y^1} \wedge \frac{dY^2}{Y^2} \,.
\end{align}
Here the $G_{ij}$ are given in terms of the $\chi_{ij}$ by the expressions 
\begin{align}\label{Ggivenchi}
G_{01} &= \cosh\rho \sinh\rho \cos\theta \, [  \sin\theta \, (\sinh \rho \, \chi_{12} +i \cosh\rho \, \chi_{02}) - i \cos\theta \, \chi_{01}], \cr
G_{02} &= - \cosh\rho \sinh\rho \sin\theta \, \left[ \cos\theta \, (\sinh \rho \, \chi_{12} +i \cosh\rho \, \chi_{02}) + i \sin\theta \, \chi_{01} \right], \cr
G_{12} &= \sinh^2\rho \cos\theta \sin\theta \, (\cosh \rho \, \chi_{12} + i \sinh \rho \, \chi_{02})\,.
\end{align}
These relations can be inverted to express the $\chi_{ij}$ as linear combinations of $G_{ij}$ since the matrix of coefficients is non-singular. The inverse relations of the 2-form expressions \eqref{EstoYs} are
\begin{eqnarray}
 \frac{dY^0}{Y^0} \wedge \frac{dY^1}{Y^1} &=& \frac{i}{\sinh\rho} \left[ \frac{1}{\cosh\rho} E^{01} - \tan\theta \, E^{02} \right] - \frac{\tan\theta}{\cosh\rho} \, E^{12}, \cr
 \frac{dY^0}{Y^0} \wedge \frac{dY^2}{Y^2} &=& \frac{i}{\sinh\rho} \left[ \frac{1}{\cosh\rho} E^{01} + \cot\theta \, E^{02} \right] + \frac{\cot\theta}{\cosh\rho} \, E^{12},\cr
 \frac{dY^1}{Y^1} \wedge \frac{dY^2}{Y^2} &=& \frac{1}{\sinh\rho \cos\theta \sin\theta} \left[ i E^{02} + \coth \rho \, E^{12} \right].
\end{eqnarray}
These combinations of $\{E^{01}, E^{02}, E^{12}\}$ have the important property that their exterior derivatives vanish. This fact will come in handy when we compute the exterior derivative of $G$. Let us now turn to solving the equation $dG=0$. The left hand side of this equation reads
\begin{multline}\label{dGeqn}
dG = dG_{01} \wedge  \left(\frac{i}{\sinh\rho} \left[ \frac{1}{\cosh\rho} {\bE}^{01} - \tan\theta \, {\bE}^{02} \right] 
- \frac{\tan\theta}{\cosh\rho} \, {\bE}^{12} \right) \cr
+ dG_{02} \wedge \left( \frac{i}{\sinh\rho} \left[ \frac{1}{\cosh\rho} {\bE}^{01} + \cot\theta \, {\bE}^{02} \right] 
+ \frac{\cot\theta}{\cosh\rho} \, {\bE}^{12} \right) \cr
+ dG_{12} \wedge \left( \frac{1}{\sinh\rho \cos\theta \sin\theta} \left[ i {\bE}^{02} + \coth \rho \, {\bE}^{12} \right] \right)\,,
\end{multline}
with 
\begin{multline}\label{dGref}
dG_{ij} = (K_0 G_{ij}) \, {\bE}^0 + (K_1 G_{ij}) \, {\bE}^1 + (K_2 G_{ij}) \, {\bE}^2 
 + (K_{\bar 0} G_{ij}) \, {\bE}^{\bar 0} + (K_{\bar 1} G_{ij}) \, {\bE}^{\bar 1} + (K_{\bar 2} G_{ij}) \, {\bE}^{\bar 2} \,,
\end{multline}
where $K_A$ is the vector field dual to the 1-form $E^A$. Their explicit forms have been collected in Appendix \ref{Kfields}. Now, \eqref{dGeqn} is an equation for a 3-form on the world-volume of the dual-giant and one should set the coefficients of the linearly independent 3-forms to zero. As before, we will do this pretending that this is a bulk 3-form equation given the form of our ansatz. Such a solution would lead to a spacetime 2-form $G$, independent of the particular polynomial that defines the dual-giant. We implement this procedure below. 

Given that the equation describing the dual-giant is holomorphic in the variables $Y^i$, it follows that ${\bE}^{012}=0$. This is true irrespective of the precise form of the defining polynomial $f(Y^i)=0$. Also, from \eqref{dGeqn}, it follows that two of the three indices in the 3-form have to be unbarred. Given these constraints, there are precisely nine independent 3-forms that appear on the right hand side of that equation if we substitute \eqref{dGref} into \eqref{dGeqn}. After some algebra we find that imposing $dG=0$ is equivalent to imposing the nine constraints
\begin{equation}
K_{\bar 0} G_{ij} = K_{\bar 1} G_{ij} = K_{\bar 2} G_{ij} = 0\quad \text{for}\quad i,j\in \{0,1,2\}\,.
\end{equation}
These make $G_{ij}$ to be functions of $Y^0, Y^1, Y^2$ and not their conjugates (refer Appendix \ref{Kfields} for details).

We can summarize our results so far as follows: any 1/8-BPS dual-giant in $AdS_5 \times S^5$ with non-trivial world-volume electromagnetic fields is specified by 
\begin{itemize}
\item a holomorphic function $$f(Y^0, Y^1, Y^2) \quad\text{and}$$
\item a holomorphic 2-form
$$G = \sum_{i,j=0,1,2}G_{ij} \frac{dY^i}{Y^i} \wedge \frac{dY^j}{Y^j}\,, $$
with $G_{ij} = G_{ij} (Y^0, Y^1, Y^2)$. 
\end{itemize}
The world-volume is obtained by taking the intersection of the zero-set of $f(Y^0, Y^1, Y^2)$ with $AdS_5 \times S^5$ and the field strength of the world-volume gauge field is given by the real part of the 2-form $G$ pulled back onto the world-volume.

This completes our general analysis of the BPS equations for dual-giants in the $AdS_5 \times S^5$  background of type IIB supergravity. Even though we restricted ourselves to $AdS_5 \times S^5$, our analysis applies to any of the known supersymmetric backgrounds of the type $AdS_5 \times X_5$ where $X_5$ is a Sasaki-Einstein manifold of \cite{Gauntlett:2004zh, Gauntlett:2004yd}. In these cases we have to replace our $e^9$ with the 1-form dual to the Reeb vector field of the corresponding $X_5$.

\section{General giant solution}

Our analysis of the previous section can be extended to the case of giant gravitons in a straightforward manner. The BPS equations with electromagnetic fluxes have already been written out in equations \eqref{bpsqnsgg} and \eqref{bpsfeqns}, we collect them below for convenience. The equations 
\begin{align}
{\bf E}^1 = {\bf E}^2 &= {\bf E}^{0\bar{0} 56} = 0\,,  \cr
{\bf E}^0 \wedge {\bf E}^A \wedge \omega &= 0\quad\text{for}\quad A = \{5,6\}\quad\text{and}\cr
\omega\wedge\omega &=0  
\end{align}
define the world-volume of the time-like giant graviton while the equations
\begin{subequations}\label{Fgiants}
\begin{align}
F\wedge {\bf E}^{0\bar{0}} &= 0 \\ 
F\wedge {\bf E}^0 \wedge \{{\bf E}^5, {\bf E}^6, {\bf E}^{\bar 5}, {\bf E}^{\bar 6} \} &= 0 \\
F\wedge F &= 0\\
F\wedge  ( {\bf E}^{5\bar{5}} +  {\bf E}^{6 \bar{6}}) &= 0 \\
F\wedge \{ {\bf E}^{56}, {\bf E}^{{\bar 5}{\bar 6}} \} &= 0 \, .
\end{align}
\end{subequations}
are the constraints on the field strength $F$. The $F$-independent equations are solved by the Mikhailov giants \cite{Mikhailov:2000ya, Kim:2006he, ashoknemani}, which are described in terms of the zero-set
\be\label{defgg}
f(X_1, X_2, X_3) = 0\,, 
\ee
with $X_i = \Phi^0\, Z_i$
\begin{align}
Z_1 &= \sin \alpha \, e^{i\xi_1}\qquad Z_2 = \cos \alpha \sin\beta \, e^{i\xi_2}\qquad Z_3 = \cos \alpha \cos\beta \, e^{i\xi_3}\cr 
\Phi^0 &= e^{i\phi_0}\qquad\text{and}\qquad \Phi^1 = \Phi^2 = 0 \,. 
\end{align}
The F-dependent BPS equations are now solved by the ansatz
\be
F = \text{Re}\big[\psi_{05}{\bE}^{05} + \psi_{06}{\bE}^{06} + \psi_{56}{\bE}^{56}\big]  \,.
\ee
Substituting this ansatz into the expression for $X^{ij}$ and simplifying the resulting expression using the BPS conditions, one finds that the dual $2$-form $\tilde{X}$ takes the form
\be
\tilde{X} = \text{Im}\big[\psi_{05}{\bE}^{05} + \psi_{06}{\bE}^{06} + \psi_{56}{\bE}^{56}\big]  \,.
\ee
The equation of motion and the Bianchi identity can therefore be combined into the single equation
\be
dG \equiv d(F+i\, \tilde{X}) = 0 \,.
\ee
Proceeding as before, we find the following result: any 1/8-BPS giant graviton in $AdS_5\times S^5$ with non-trivial world-volume electromagnetic fields is specified by a polynomial equation of the form \eqref{defgg} along with a complex 2-form
\be
G = \sum_{i=1,2,3}G_{ij} \frac{dX_i}{X_i}\wedge \frac{dX_j}{X_j} \,,
\ee
with $G_{ij} = G_{ij}(X_1, X_2, X_3)\,$. The world-volume is obtained by taking the intersection of $f(X_i) =0$ with $AdS_5\times S^5$, while the field strength is given by the real part of the $2$-form G pulled back onto the world-volume. 

\section{Explicit examples}

Finally we would like to apply our results to recover the known BPS giants with electromagnetic waves. These are EM waves on the round $S^3 \subset AdS_5$ dual-giant \cite{Sinha:2007ni} and the round $S^3 \subset S^5$ giant \cite{Kim:2005mw}. In the following we demonstrate how to recover the BPS electromagnetic fields on the round dual-giant. 
 
Recall that the half-BPS round $S^3$ dual-giant \cite{Grisaru:2000zn, Hashimoto:2000zp} is described by the intersection of $\{ Y^0-c = 0, Z_2 = Z_3 =0 \}$ in ${\mathbb C}^{1,2} \times {\mathbb C}^3$ with $AdS_5 \times S^5$. It is easy to check that one has 
\begin{equation}
\label{dggeq1}
i {\bf E}^0 + \tanh \rho \, {\bf E}^1 = 0.
\end{equation}
This implies that ${\bf E}^{01} = 0$ on the world-volume. The 2-form $G$ for this example simplifies to
\begin{eqnarray}
\label{dggeq2}
G &=& \chi_{02} {\bf E}^{02} + \chi_{12} {\bf E}^{12} \cr
&=&  G_{12} (Y^0 = c, Y^1, Y^2)\frac{1}{\sinh\rho \cos\theta \sin\theta} \left[ i {\bE}^{02} + \coth \rho \, {\bE}^{12} \right]\,,
\end{eqnarray}
where, in going from the first line to the second line one uses (\ref{dggeq1}) and (\ref{Ggivenchi}). To compare with \cite{Sinha:2007ni} we need to pick a gauge to fix the world-volume diffeomorphisms. We choose\footnote{This is a different gauge from the one chosen in \cite{Sinha:2007ni} and is related to it by a boost. We have checked that if we choose the standard static gauge $t = \tau, \theta = \sigma_1, \phi_1 = \sigma_2, \phi_2 = \sigma_3$ our general solution for $F$ does reproduce the answer obtained in \cite{Sinha:2007ni}.}  
\begin{equation}\label{boostedansatz}
\phi_0 = \tau, ~~ \theta = \sigma_1, ~~ \phi_1 = \tau + \sigma_2, ~~ \phi_2 = \tau + \sigma_3, 
\end{equation}
and the ansatz
\begin{equation}
\rho = \rho_0, ~~ \xi_1 = -\tau + \xi_1^{(0)}, ~~ \alpha = \frac{\pi}{2}
\end{equation}
which solves the BPS equations that do not involve the gauge field. Substituting this solution into (\ref{dggeq2}) we obtain
\begin{equation}
\label{dggeg3}
G = - G_{12} (Y^0, Y^1, Y^2 ) \,  \left[ d\sigma_2 \wedge d\sigma_3 +i \,  d\sigma_1 \wedge (\cot \sigma_1 \, d\sigma_2 + \tan \sigma_1 \, d\sigma_3) \right]\,,
\end{equation}
where $Y^0 (= c) = \cosh\rho_0 \, e^{i\xi_1^{(0)}}$, $Y^1= \sinh\rho_0 \, \cos\sigma_1 e^{i(\sigma_2 + \xi_1^{(0)})}$ and $Y^2 = \sinh\rho_0 \, \sin\sigma_1 e^{i(\sigma_3 + \xi_1^{(0)})}$. The real part of $G$ gives the BPS solutions for $F$.
 
On the other hand one can carry out a direct analysis of the problem of finding the BPS gauge field configurations on the $S^3$ dual-giant similar to \cite{Sinha:2007ni}. The BPS constraints on $F$ result in the equations
\be
F_{0i} = 0 \qquad\text{and} \qquad F_{13} = \tan^2\sigma_1\, F_{12} \,. 
\ee 
Thus there are two independent components of $F$, $F_{12}$ and $F_{23}$. One can check that if we define the complex combination
\be
{\cal G} = F_{23} - i \tan\sigma_1 F_{12}   \,,
\ee
the equation of motion and the Bianchi identity can be combined into the equations
\begin{align}
\p_0\, {\cal G} = 0 \quad\text{and}\quad
\left[\partial_{\sigma_1} - i (\tan\sigma_1 \partial_{\sigma_2} - \cot\sigma_1 \partial_{\sigma_3})\right]{\cal G} = 0 \,.
\end{align}
The first equation implies that the components $F_{23}$ and $F_{12}$ (and therefore $F_{13}$) are $\tau$-independent. The solution to the second equation is then given by
\be
{\cal G}(\sigma_i) = \sum_{m, n} C_{m, n}(\cos\sigma_1 e^{i\sigma_2})^{m}(\sin\sigma_1 e^{i\sigma_3})^{n} := {\cal G} (\cos\sigma_1 \, e^{i\sigma_2}, \sin\sigma_1 \, e^{i\sigma_3})\,.
\ee
Identifying ${\cal G} = -G_{12}$ one finds that this is precisely the solution we have obtained in (\ref{dggeg3}). 

The analogous analysis for the round giant can be carried out along similar lines. We will not do this explicitly here. 

\section{Discussion}

We have constructed all the dual-giant and giant graviton solutions of a D3-brane with all its bosonic world-volume fields turned on, and which propagate in the $AdS_5 \times S^5$ background of type IIB string theory. The simplifications that occur in solving for these configurations can only be attributed to the magic of supersymmetry. We have, however, assumed that the world-volume gauge field strength $F$ is given by the restriction of a bulk 2-form. One could suspect that there may be other solutions which may not satisfy our assumption about the gauge field and it will be important to check if there are solutions for $F$ outside the realm of our assumption.

We have demonstrated that all the simple known examples of BPS EM waves on giants can be recovered from our construction. From this analysis, it becomes clear that our solutions provide a world-volume reparametrization independent description of the supersymmetric field strength on the D-brane. 

The solutions we found involve a holomorphic 2-form whose real part restricted to the world-volume gives rise to the gauge field strength. This 2-form has three independent holomorphic functions in its specification. However, when we pull this back onto the world-volume of the D3-brane, we do not expect all the three functions to play independent roles. Since one of $\{{\bf E}^0, {\bf E}^1, {\bf E}^2\}$ may be eliminated in favour of the other two, there is a unique independent complex 2-form and consequently, only one independent coefficient function, determined by the $dG=0$ condition. 

The techniques developed here should be easily applicable in other contexts. For instance, it would be interesting to construct all the giant and dual-giant like configurations of M2 and M5 branes along with their world-volume fluxes propagating in the $AdS_4 \times S^7$ and $AdS_7 \times S^4$ backgrounds of M-theory \cite{Mikhailov:2000ya, Bhattacharyya:2007sa}. 
It is also important to construct classical BPS solutions that involve turning on the fermionic fields on the world-volume of the branes as one expects such solutions to exist. 

One anticipates that there would be several applications of our solutions since giant gravitons have so far played an important role in furthering our understanding of AdS/CFT. 
For instance, the quantization of the parameter spaces of giant and dual-giant solutions has lead to the verification of the AdS/CFT duality in specific subsectors that preserve (at least) four supercharges \cite{Suryanarayana:2004ig, Biswas:2006tj, Mandal:2006tk, Martelli:2006vh, Basu:2006id, ashoknemani}. A complete verification of AdS/CFT in such 1/8-BPS sectors would require the knowledge of {\it all} BPS states that preserve a specific set of supersymmetries. In order to achieve this, one would have to obtain the most general supersymmetric D3-brane configuration, with both bosonic as well as fermionic world-volume fields turned on. Following the strategy used in \cite{Biswas:2006tj, ashoknemani}, one could then characterize the solution space of the most general 1/8-BPS giant or dual-giant configuration in an appropriate way and do a geometric quantization of the resulting configuration space \cite{Das:2000fu, Suryanarayana:2004ig, Mandal:2005wv, Biswas:2006tj, Mandal:2006tk, Martelli:2006vh, Basu:2006id, ashoknemani}. Such an exercise will help complete the programme of counting BPS states in the 1/8-BPS supersymmetric sectors and their comparison with the corresponding gauge theory answers \cite{Kinney:2005ej}. We hope to address some of these issues in the future.

\newpage

\begin{appendix}

\section{Simplifying the determinant}
Here we give some relevant details of the manipulations we perform towards simplifying the determinant $\det (h+F)$. Recall that $h_{ij} = \mathfrak{e}^a_i \mathfrak{e}^b_j \eta_{ab}$. Using the definition of the determinant of a $4 \times 4$ matrix 
\begin{align}
\det (h+F) &= \frac{1}{4!} \epsilon^{ijkl} \epsilon^{mnpq} (h_{im} + F_{im}) \cdots (h_{lq} + F_{lq}) \cr
&= \det h + \det F + \frac{1}{4} \epsilon^{ijkl} \epsilon^{mnpq} h_{im} h_{jn} F_{kp} F_{lq} \cr
&= \det h + \det F + \frac{1}{8} (\epsilon^{ijkl} \mathfrak{e}^{a_1}_i \mathfrak{e}^{a_2}_j F_{kl}) (\epsilon^{mnpq} \mathfrak{e}^{b_1}_m \mathfrak{e}^{b_2}_n F_{pq}) \, \eta_{a_1 b_1} \eta_{a_2 b_2} \cr
&:= \det h + \det F - (\mathfrak{e}^{09} \wedge F)^2 + \sum_{A=1} \left[|\mathfrak{e}^9 \wedge {\bf E}^A \wedge F|^2 - |\mathfrak{e}^0 \wedge {\bf E}^a \wedge F|^2 \right] + \sum_{a<b} (\mathfrak{e}^{ab} \wedge F)^2 \,. \cr
\end{align}
In going from the second line to the third line above we have used the identity
\begin{equation}
h_{im} h_{jn} \epsilon^{ijkl} \epsilon^{mnpq} (F_{kp} F_{lq} + F_{kl} F_{qp} + F_{kq} F_{pl} ) = 0 \,.
\end{equation}
In the third line we use the notation
\begin{equation}
\mathfrak{e}^{ab} \wedge F := \frac{1}{2} \epsilon^{ijkl} \mathfrak{e}^a_i \mathfrak{e}^b_j F_{kl}.
\end{equation}
Next notice that 
\begin{align}
\det h &= \frac{1}{4!} (\epsilon^{ijkl} \mathfrak{e}^{a_1}_i \mathfrak{e}^{a_2}_j  \mathfrak{e}^{a_3}_k \mathfrak{e}^{a_4}_l) (\epsilon^{mnpq} \mathfrak{e}^{b_1}_m \mathfrak{e}^{b_2}_n \mathfrak{e}^{b_3}_p \mathfrak{e}^{b_4}_q) \eta_{a_1b_1} \cdots \eta_{a_4 b_4} \, ,\cr
\det F &= ({\rm Pf} [F])^2 
\end{align}
where ${\rm Pf} [F] = \frac{1}{8} \epsilon^{ijkl} F_{ij} F_{kl}$. It has been shown in \cite{ashoknemani} that $\det h$, written in terms of the $E^A$, reads
\begin{multline}
\det h = - \sum_{A<B} | \mathfrak{e}^{09} \wedge {\bf E}^{AB} |^2 - (\mathfrak{e}^{09} \wedge \Omega)^2 \cr
- \sum_{A<B<C} \left[|\mathfrak{e}^0 \wedge {\bf E}^{ABC}|^2 -|\mathfrak{e}^9 \wedge {\bf E}^{ABC}|^2\right]- \sum_A \left[|\mathfrak{e}^0 \wedge {\bf E}^A \wedge \Omega|^2 - |\mathfrak{e}^9 \wedge {\bf E}^A \wedge \Omega|^2 \right] \cr
+ |{\bf E}^{1256}|^2 + \sum_{A<B} |{\bf E}^{AB} \wedge \Omega|^2 + \frac{1}{4} (\Omega \wedge \Omega)^2
\end{multline}
where $\Omega = \tilde \omega - \omega = -\frac{i}{2} \sum_A {\bf E}^{A\bar A}$. In showing this one makes repeated use of the identity
\begin{equation}
\mathfrak{e}^{abc[d} \mathfrak{e}^{efgh]} = 0\,. 
\end{equation}
Finally one can show, by rewriting $\mathfrak{e}^a_i$ in terms of ${\bf E}^A_i$, that 
\begin{equation}
\sum_{a<b=1}^8 (\mathfrak{e}^{ab} \wedge F)^2 = \sum_{A<B} |{\bf E}^{AB} \wedge F|^2 + (\Omega \wedge F)^2 - (\Omega \wedge \Omega) \, {\rm Pf} [F] \,,
\end{equation}
where we used the following identity
\begin{equation}
F_{kl} F_{pq} \, \epsilon^{ijkl} \epsilon^{mnpq}  \mathfrak{e}^a_i (\mathfrak{e}^b_j \mathfrak{e}^c_m \mathfrak{e}^d_n + \mathfrak{e}^d_j \mathfrak{e}^b_m \mathfrak{e}^c_n + \mathfrak{e}^c_j \mathfrak{e}^d_m \mathfrak{e}^b_n) = 4 \, {\rm Pf} (F) \, \epsilon^{ijkl} \mathfrak{e}^a_i \mathfrak{e}^b_j \mathfrak{e}^c_k \mathfrak{e}^d_l \, .
\end{equation}

\section{Vector fields and holomorphic functions}\label{Kfields}

Consider a completely general function of all the coordinates in $AdS_5\times S^5$:
\begin{equation}
f(r,\theta, \alpha, \beta, \phi_0, \phi_1, \phi_2, \xi_1, \xi_2, \xi_3) = 0 \,.
\end{equation}
Suppose this is one of the defining equations of a D3-brane world-volume. This leads to a differential constraint on the 1-forms in spacetime when pulled back onto the world-volume:
\begin{align}
{\rm P}\left[f_r\, dr + f_{\theta}\, d\theta + f_{\alpha}\, d\alpha + f_{\beta}\, d\beta + \sum_{i=0,1,2} f_{\phi_i}\, d\phi_i + \sum_{i=1,2,3}f_{\xi_i}\, d\xi_i\right] = 0 \,,
\end{align}
where $f_x = \p_x f$. It is possible to rewrite each of these 1-forms in terms of the complex 1-forms \eqref{cplxforms} using the explicit frames used in \eqref{adsframe} and \eqref{sframe}. This leads to the differential constraint 
\begin{multline}\label{dFeqn}
\left[ f_{\rho} -  i\, \big(\tanh\rho\, f_{\phi_0} + \coth\rho\, (f_{\phi_1}+f_{\phi_2})\big) \right]\, {\bf E}^1 + 
\left[ f_{\rho} + i\, \big(\tanh\rho\, f_{\phi_0} + \coth\rho\, (f_{\phi_1}+f_{\phi_2})\big) \right]\, \overline{\bf E^1}\cr
+ \frac{1}{\sinh\rho}\left[ f_{\theta}+i\, \big(\tan\theta\, f_{\phi_1}-\cot\theta\, f_{\phi_2} \big)\right]\, {\bf E}^2 +
\frac{1}{\sinh\rho}\left[ f_{\theta}-i\, \big(\tan\theta\, f_{\phi_1}-\cot\theta\, f_{\phi_2} \big)\right]\, 
\overline{\bf E^2}\cr
+\left[f_{\alpha}-i \big(\cot\alpha\, f_{\xi_1} - \tan\alpha(f_{\xi_2}+f_{\xi_3}) \big) \right]{\bf E}^5
+\left[f_{\alpha}+i\big(\cot\alpha\, f_{\xi_1} - \tan\alpha(f_{\xi_2}+f_{\xi_3}) \big) \right]\overline{\bf E^5}\cr
+\frac{1}{\cos\alpha}\left[f_{\beta}-i\big(\cot\beta\, f_{\xi_2}-\tan\beta\, f_{\xi_3})\big)\right]{\bf E}^6 
+\frac{1}{\cos\alpha}\left[f_{\beta}+i\big(\cot\beta\, f_{\xi_2}-\tan\beta\, f_{\xi_3})\big)\right]\overline{\bf E^6}\cr
 +\left[\sum_{i=0,1,2} f_{\phi_i} + \sum_{i=1,2,3} f_{\xi_i}\right]\big({\mathfrak e}^0+{\mathfrak e}^9\big)+\left[\sum_{i=0,1,2}f_{\phi_i} - \sum_{i=1,2,3} f_{\xi_i}\right]\big({\mathfrak e}^0-{\mathfrak e}^9\big) =0
\,.
\end{multline}
For dual-giants which satisfy ${\bE}^5 = {\bE}^6 = 0$, this equation is compactly written in the form
\be
\sum_{A=0}^{2} \left[(K_A f)\,  {\bf E}^A + (K_{\bar A}\, f)\, {\bf E}^{\bar A}\right] = 0 \,,
\ee
where the $K$s are the vector fields written out explicitly in \eqref{dFeqn}. In \cite{ashoknemani} it was shown that the wobbling dual-giants are described by functions that satisfy the constraints
\be
K_{\bar 0}f = K_{\bar 1}f = K_{\bar 2} f = 0 \,.
\ee
This follows from the BPS equations. The solutions to these equations are functions that depend purely on the variables $Y^i$, where
\be
Y^i = \Phi^i\, Z_1 \quad \text{and}\quad Z_2=Z_3 = 0\,.
\ee
Here, the $(\Phi^0, \Phi^1, \Phi^2) \in \mathbb{C}^{1,2}$ and $(Z_1, Z_2, Z_3)\in \mathbb{C}^3$ are coordinates on $AdS_5$ and $S^5$ respectively since they satisfy $|\Phi^0|^2 - |\Phi^1|^2 - |\Phi^2|^2 = 1$ and $|Z_1|^2 + |Z_2|^2 + |Z_3|^2 = 1$.   

Analogously, giant gravitons are those for which ${\bE}^1 = {\bE}^2 = 0$ as a result of which the constraint on the 1-forms takes the form
\be
\sum_{a=0,5,6}\left[(K_A\, f)\, {\bf E}^A + (K_{\bar A}\, f)\, {\bf E}^{\bar A}\right] = 0 \,. 
\ee
The BPS equations require the function $f$ that describes the world-volume of a giant graviton to satisfy 
\be
K_{\bar 0} f = K_{\bar 5} f = K_{\bar 6} f = 0 \,.
\ee
This makes the function to depend purely on the variables $X_i$, where 
\be
X_i = Z_i\, \Phi^0 \quad \text{and}\quad \Phi_1 = \Phi_2 = 0 \,.
\ee
These are the equations that describe the Mikhailov giants. 

\end{appendix}


\begin{thebibliography}{99}

\bibitem{McGreevy:2000cw}
  J.~McGreevy, L.~Susskind and N.~Toumbas,
  ``Invasion of the giant gravitons from anti-de Sitter space,''
  JHEP {\bf 0006}, 008 (2000)
  [arXiv:hep-th/0003075].

\bibitem{Grisaru:2000zn}
  M.~T.~Grisaru, R.~C.~Myers and O.~Tafjord,
  ``SUSY and Goliath,''
  JHEP {\bf 0008}, 040 (2000)
  [arXiv:hep-th/0008015].

\bibitem{Hashimoto:2000zp}
  A.~Hashimoto, S.~Hirano and N.~Itzhaki,
  ``Large branes in AdS and their field theory dual,''
  JHEP {\bf 0008}, 051 (2000)
  [arXiv:hep-th/0008016].

\bibitem{Mikhailov:2000ya}
  A.~Mikhailov,
  ``Giant gravitons from holomorphic surfaces,''
  JHEP {\bf 0011}, 027 (2000)
  [arXiv:hep-th/0010206].

\bibitem{Kim:2006he}
  S.~Kim and K.~M.~Lee,
  ``1/16-BPS black holes and giant gravitons in the AdS(5) x S**5 space,''
  JHEP {\bf 0612}, 077 (2006)
  [arXiv:hep-th/0607085].

\bibitem{ashoknemani}
  S.~K.~Ashok and N.~V.~Suryanarayana,
  ``Counting Wobbling Dual-Giants,''
  JHEP {\bf 0905}, 090 (2009)
  [arXiv:0808.2042 [hep-th]].

\bibitem{Kim:2005mw}
  S.~Kim and K.~M.~Lee,
  ``BPS electromagnetic waves on giant gravitons,''
  JHEP {\bf 0510}, 111 (2005)
  [arXiv:hep-th/0502007].
  
\bibitem{Sinha:2007ni}
  A.~Sinha and J.~Sonner,
  ``Black Hole Giants,''
  JHEP {\bf 0708}, 006 (2007)
  [arXiv:0705.0373 [hep-th]].

\bibitem{Mandal:2006tk}
  G.~Mandal and N.~V.~Suryanarayana,
 ``Counting 1/8-BPS dual-giants,''
  JHEP {\bf 0703}, 031 (2007)
  [arXiv:hep-th/0606088].

\bibitem{Cederwall:1996pv}
  M.~Cederwall, A.~von Gussich, B.~E.~W.~Nilsson and A.~Westerberg,
  ``The Dirichlet super-three-brane in ten-dimensional type IIB
  supergravity,''
  Nucl.\ Phys.\  B {\bf 490}, 163 (1997)
  [arXiv:hep-th/9610148].

\bibitem{Bergshoeff:1996tu}
  E.~Bergshoeff and P.~K.~Townsend,
``Super D-branes,''
  Nucl.\ Phys.\  B {\bf 490}, 145 (1997)
  [arXiv:hep-th/9611173].
  
\bibitem{Gauntlett:2004zh}
  J.~P.~Gauntlett, D.~Martelli, J.~Sparks and D.~Waldram,
``Supersymmetric AdS(5) solutions of M-theory,''
  Class.\ Quant.\ Grav.\  {\bf 21}, 4335 (2004)
  [arXiv:hep-th/0402153].

\bibitem{Gauntlett:2004yd}
  J.~P.~Gauntlett, D.~Martelli, J.~Sparks and D.~Waldram,
``Sasaki-Einstein metrics on S(2) x S(3),''
  Adv.\ Theor.\ Math.\ Phys.\  {\bf 8}, 711 (2004)
  [arXiv:hep-th/0403002].

\bibitem{Bhattacharyya:2007sa}
  S.~Bhattacharyya and S.~Minwalla,
  ``Supersymmetric states in M5/M2 CFTs,''
  JHEP {\bf 0712}, 004 (2007)
  [arXiv:hep-th/0702069].
  
\bibitem{Suryanarayana:2004ig}
  N.~V.~Suryanarayana,
  ``Half-BPS giants, free fermions and microstates of superstars,''
  JHEP {\bf 0601}, 082 (2006)
  [arXiv:hep-th/0411145].

\bibitem{Biswas:2006tj}
  I.~Biswas, D.~Gaiotto, S.~Lahiri and S.~Minwalla,
  ``Supersymmetric states of N = 4 Yang-Mills from giant gravitons,''
  JHEP {\bf 0712}, 006 (2007)
  [arXiv:hep-th/0606087].

\bibitem{Martelli:2006vh}
  D.~Martelli and J.~Sparks,
  ``Dual giant gravitons in Sasaki-Einstein backgrounds,''
  Nucl.\ Phys.\  B {\bf 759}, 292 (2006)
  [arXiv:hep-th/0608060].

\bibitem{Basu:2006id}
  A.~Basu and G.~Mandal,
  ``Dual giant gravitons in AdS(m) x Y**n (Sasaki-Einstein),''
  JHEP {\bf 0707}, 014 (2007)
  [arXiv:hep-th/0608093].

\bibitem{Das:2000fu}
  S.~R.~Das, A.~Jevicki and S.~D.~Mathur,
  ``Giant gravitons, BPS bounds and noncommutativity,''
  Phys.\ Rev.\  D {\bf 63}, 044001 (2001)
  [arXiv:hep-th/0008088].

\bibitem{Mandal:2005wv}
  G.~Mandal,
  ``Fermions from half-BPS supergravity,''
  JHEP {\bf 0508}, 052 (2005)
  [arXiv:hep-th/0502104].

\bibitem{Kinney:2005ej}
  J.~Kinney, J.~M.~Maldacena, S.~Minwalla and S.~Raju,
 ``An index for 4 dimensional super conformal theories,''
  Commun.\ Math.\ Phys.\  {\bf 275}, 209 (2007)
  [arXiv:hep-th/0510251].

 \end{thebibliography}
\end{document}